\documentclass[preprint,preprintnumbers, prd, floatfix,  superscriptaddress,
nofootinbib] {revtex4}
\usepackage{graphicx}
\usepackage{epsfig}
\usepackage{bm}
\usepackage{amssymb}
\usepackage{float}
\usepackage{amsmath}
\usepackage{dcolumn}
\usepackage{cancel}
\usepackage[colorlinks]{hyperref}
\usepackage[usenames,dvipsnames]{color}
\usepackage{epstopdf}
\hypersetup{
     breaklinks=true,
    pdfstartview={FitH},    colorlinks=true,       % false: boxed links; true: colored links
    linkcolor=blue,          % color of internal links
    citecolor=red,        % color of links to bibliography
    filecolor=magenta,      % color of file links
    urlcolor=blue,           % color of external links
    anchorcolor=green,      % Color for anchor text
    linktocpage=true
}

\def\doi{http://doi.org}

\begin{document}

 \title{Running cosmological constant with observational tests}

 \author{Chao-Qiang Geng}
\email{geng@phys.nthu.edu.tw}
\affiliation{Chongqing University of Posts \& Telecommunications, Chongqing, 400065, 
China}
 \affiliation{National Center for Theoretical Sciences, Hsinchu,
Taiwan 300}
\affiliation{Department of Physics, National Tsing Hua University,
Hsinchu, Taiwan 300}

\author{Chung-Chi Lee}
\email{chungchi@mx.nthu.edu.tw}
 \affiliation{National Center for Theoretical Sciences, Hsinchu,
Taiwan 300}

\author{Kaituo Zhang}
\email{ktzhang@ahnu.edu.cn}
\affiliation{Anhui Normal University, Wuhu, Anhui, 24100, China}
 \affiliation{National Center for Theoretical Sciences, Hsinchu,
Taiwan 300}

\begin{abstract}

We investigate the running cosmological constant model with dark energy linearly proportional to the Hubble parameter, 
$\Lambda = \sigma H + \Lambda_0$, in which the $\Lambda$CDM limit is recovered by taking $\sigma=0$.
We derive the linear perturbation equations of gravity under the Friedmann-Lema\"itre-Robertson-Walker cosmology, 
and show the power spectra of the CMB temperature and matter density distribution.
By using the Markov chain Monte Carlo method, we fit the model to the current observational data and  
find that $\sigma H_0/ \Lambda_0 \lesssim 2.63 \times 10^{-2}$ and $6.74 \times 10^{-2}$ for $\Lambda(t)$ coupled to 
matter and radiation-matter, respectively, along with constraints on other cosmological parameters.

\end{abstract}

\maketitle

\section{Introduction} \label{sec:introduction}

The type-Ia supernova observations~\cite{Riess:1998cb,Perlmutter:1998np} have shown that our universe is undergoing a late-time accelerating expansion, which is caused by Dark Energy~\cite{Copeland:2006wr}.
The simplest way to realize such a late-time accelerating mechanism is to introduce a cosmological constant to the gravitational theory, 
such as that in the $\Lambda$CDM model.
This model fits current cosmological observations very well, but exists several difficulties, such as the ``fine-tuning"~\cite{Review1, WBook} and ``coincidence''~\cite{CC} problems.

In this work, we will concentrate on the latter problem~\cite{CP}, which has been extensively explored in the literature.
%~\cite{xxx}.
One of the popular  attempts is the running $\Lambda$ model, in which the cosmological constant evolves in time and decays to matter in the evolution of the universe~\cite{Ozer:1985ws, Carvalho:1991ut, Lima:1994gi, Lima:1995ea, Overduin:1998zv, Carneiro:2004iz, Shapiro:2009dh, Geng:2016epb, Bauer:2005rpa, Shapiro:2004ch, Dymnikova:2001ga, Alcaniz:2005dg, Barrow:2006hia}, so that the present energy densities of dark energy and dark matter are of the same order of magnitude.
Its observational applications have been investigated in Ref.~\cite{EspanaBonet:2003vk, Tamayo:2015qla, Sola:2016vis}.
In our study, we are interested in the specific model with $\Lambda = \sigma H$~\cite{Borges:2005qs, Borges:2007bh, Carneiro:2007bf, Borges:2008ii, Zimdahl:2011ae, Alcaniz:2012mh}, which would originate from the theory with the QCD vacuum condensation associated with the chiral phase transition~\cite{Schutzhold:2002pr, Klinkhamer:2009nn, Banerjee:2003fg, Ohta:2010in, Cai:2010uf}.
In this scenario, the cosmological constant decays to  matter (non-relativistic) and radiation (relativistic), leading to a large number of particles created in the cosmological evolution.
Without loss of generality, we phenomenologically extend this model to include that $\Lambda$
additionally couples to radiation with $\Lambda = \sigma H + \Lambda_0$~\cite{Basilakos:2009wi, Costa:2012xw, Gomez-Valent:2014rxa}, in which the $\Lambda$CDM limit can be realized if  $\sigma=0$.
In this scenario, when dark energy dominates the universe, the decay rate of $\Lambda$ is reduced, 
and the late-time accelerating phase occurs,
describing  perfectly the evolution history of the universe.
As a result, it is reasonable to go further to analyze the cosmological behavior of this model at the sub-horizon scale.

In this paper, we examine the matter power spectrum $P(k)$ and CMB temperature perturbations in 
the linear perturbation theory of gravity. By using the Markov chain Monte Carlo (MCMC) method, 
we perform the global fit from the current observational data and constrain the model.

This paper is organized as follows:
In Sec.~\ref{sec:model}, we introduce the $\Lambda(t)$CDM model and review its background cosmological evolutions.
In Sec.~\ref{sec:perturbation}, we calculate the linear perturbation theory and illustrate the power spectra of the matter distribution and CMB temperature by the {\bf CAMB} program~\cite{Lewis:1999bs}.
In Sec.~\ref{sec:constraints}, we use the {\bf CosmoMC} package~\cite{Lewis:2002ah} to fit the model from the observational data and show the constraints on cosmological parameters.
Our conclusions are presented  in Sec.~\ref{sec:conclusion}.

\section{The running cosmological constant model}
\label{sec:model}

We start with the Einstein equation,
% of the running cosmological constant model
given by,
\begin{eqnarray}
\label{eq:fieldeq}
R_{\mu \nu} - \frac{g_{\mu \nu}}{2}R + \Lambda(t) g_{\mu \nu} = \kappa^2 T_{\mu \nu}^M \,,
\end{eqnarray}
where $\kappa^2 = 8 \pi G$, $R=g^{\mu \nu} R_{\mu \nu}$ is the Ricci scalar, $\Lambda(t)$ is the time-dependent cosmological constant, and $T_{\mu \nu}^M$ is the energy-momentum tensor of matter and radiation.
In the Friedmann-Lema\"itre-Robertson-Walker (FLRW) case,
\begin{eqnarray}
\label{eq:pert_metric0}
ds^2 = a^2(\tau) \left[ -d\tau^2 + \delta_{ij} dx^i dx^j \right] \,,
\end{eqnarray}
we obtain,
\begin{eqnarray}
\label{eq:Friedmann-1}
&& H^2= \frac{a^2\kappa^2}{3} \left( \rho_{M} + \rho_{\Lambda} \right) \,, \\
\label{eq:Friedmann-2}
&& \dot{H} = - \frac{a^2\kappa^2}{2} \left(  \rho_{M}+  P_{M} + \rho_{\Lambda} + P_{\Lambda} \right) \,,
\end{eqnarray}
where $\tau$ is the conformal time, $H=da/(a d \tau)$ represents the Hubble parameter, $\rho_{M}$ ($P_{M}$) corresponds to the energy density (pressure) of matter and radiation, and $\rho_{\Lambda}$ ($P_{\Lambda}$) is the energy density (pressure) of the cosmological constant.
We note that from the relation of $\rho_{\Lambda} = -P_{\Lambda} = \kappa^{-2} \Lambda(t)$, derived from Eq.~(\ref{eq:fieldeq}), one has the equation of state (EoS) of $\Lambda$ to be
\begin{eqnarray}
\label{eq:wde}
w_{\Lambda} \equiv \frac{P_{\Lambda}}{\rho_{\Lambda}}= -1 \,.
\end{eqnarray}
In Eq.~(\ref{eq:fieldeq}), we consider $\Lambda(t)$ to be a linear function of the Hubble parameter, given by~\cite{Basilakos:2009wi, Costa:2012xw, Gomez-Valent:2014rxa, Alcaniz:2012mh}
\begin{eqnarray}
\label{eq:rnlam}
\Lambda = \sigma H + \Lambda_0 \,,
\end{eqnarray}
where $\sigma$ and $\Lambda_0$ are two free parameters.
From Eq.~(\ref{eq:rnlam}), we can write $\rho_{\Lambda}$ with two dimensionless parameters $\lambda_{0,1}$ as,
\begin{eqnarray}
\label{eq:rhol}
\rho_{\Lambda} = \rho_{\Lambda}^0 \left[ \lambda_0 + \lambda_1 \left( \frac{H}{H_0} \right) \right] \,,
\end{eqnarray}
where $\rho_{\Lambda}^0 \equiv \rho_{\Lambda} \lvert_{z=0}$ is the current dark energy density with the condition $\lambda_0 + \lambda_1 =1$ and $\lambda_1=\sigma H_0/(\sigma H_0+\Lambda_0)$.
Note that $\lambda_0$ has been treated as a constant of integration and set to zero in Ref.~\cite{Alcaniz:2012mh}.
Without loss of generality, we will keep $\lambda_0$ as a free parameter with the $\Lambda$CDM model recovered when $\lambda_0 \rightarrow 1$.

\begin{figure}
\centering
\includegraphics[width=0.49 \linewidth]{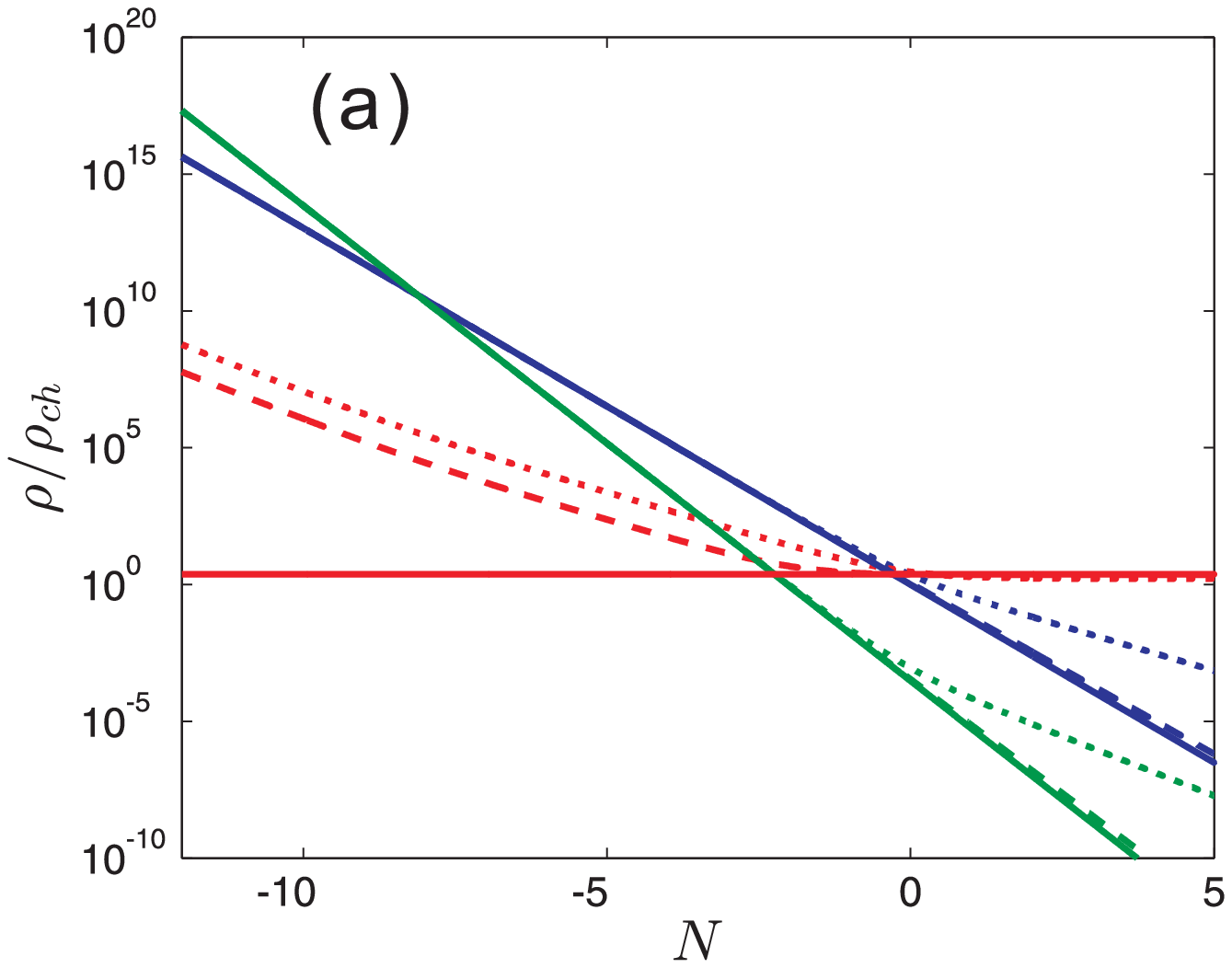}
\includegraphics[width=0.49 \linewidth]{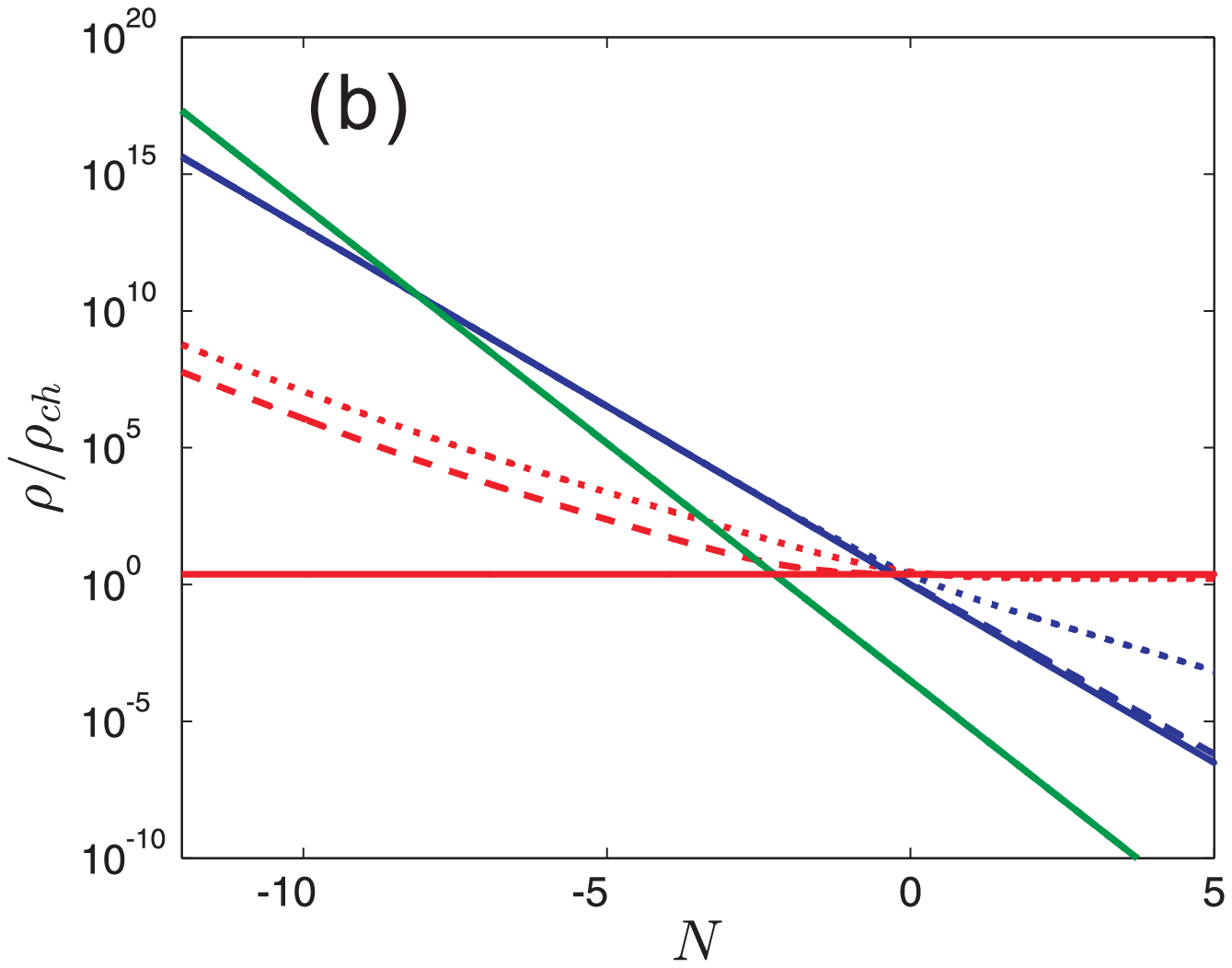}
\\
\includegraphics[width=0.49 \linewidth]{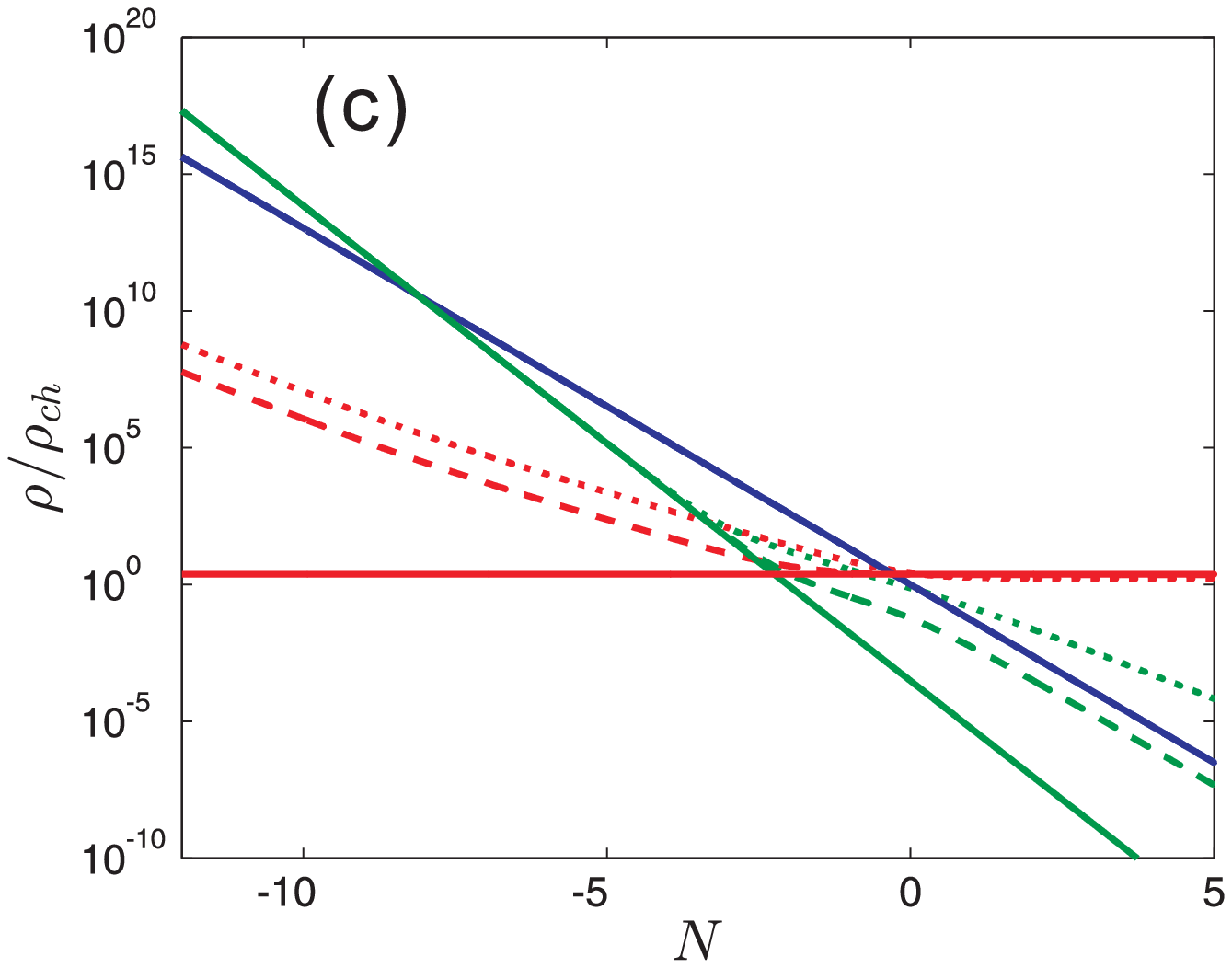}
\caption{Evolutions of $\rho_m$ (blue line), $\rho_r$ (green line) and $\rho_{\Lambda}$ (red line) with (a), (b) and (c) corresponding to $(C_r,C_m)=(1,1)$, $(0,1)$ and $(1,0)$, where the solid, dashed and dotted lines represent $(\lambda_0,\lambda_1)=(1,0)$, $(0.9,0.1)$ and $(0,1)$, respectively. 
The initial conditions are taken as $\rho_m a^{3}/\rho_{ch} = 1$ and $\rho_r a^{4}/\rho_{ch} = 3 \times 10^{-4}$ at $N \equiv \ln a = -12$,  where $\rho_{ch}$ is the characteristic energy density.}
\label{fg:1}
\end{figure}
\begin{figure}
\centering
\includegraphics[width=0.49 \linewidth]{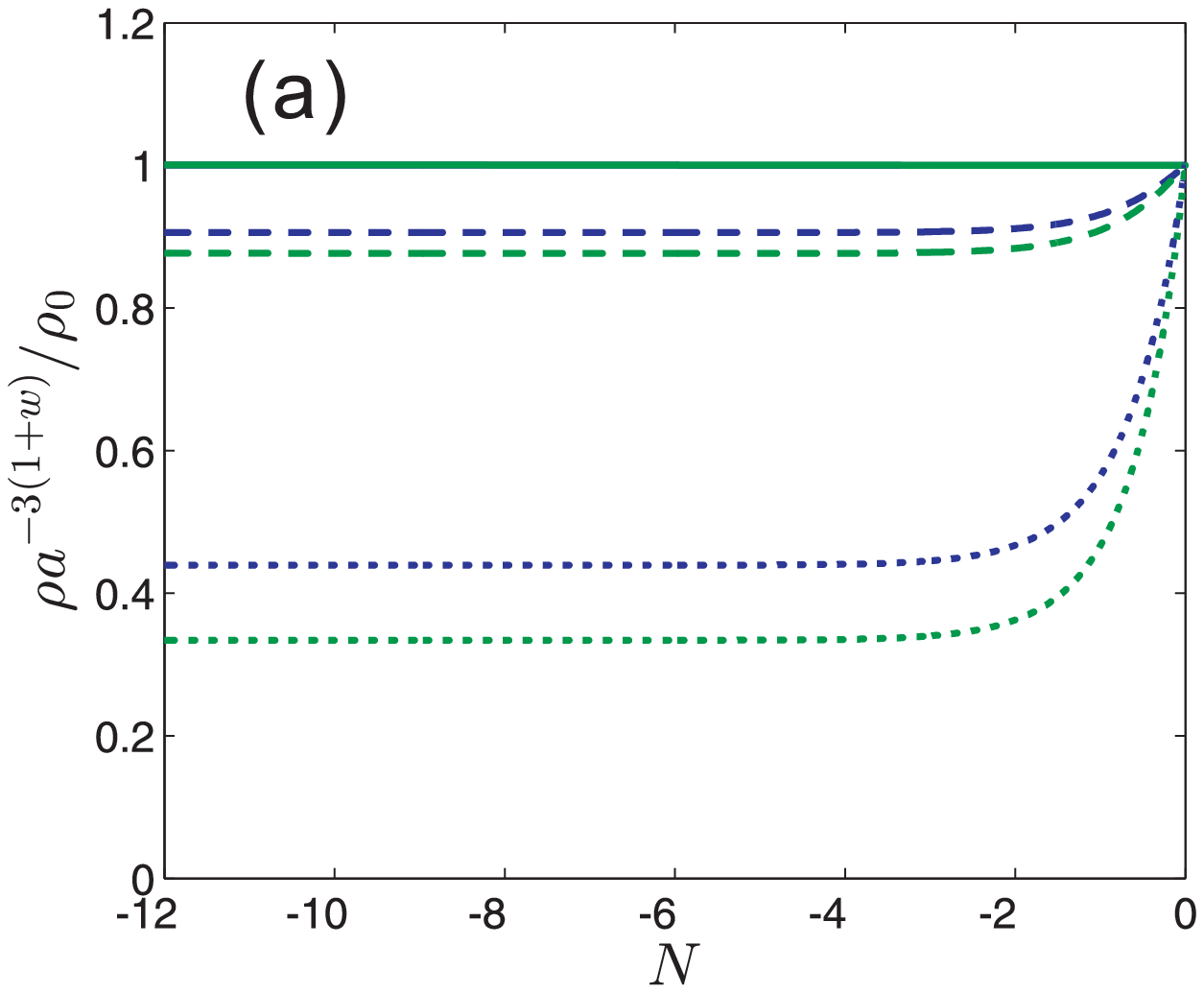}
\includegraphics[width=0.49 \linewidth]{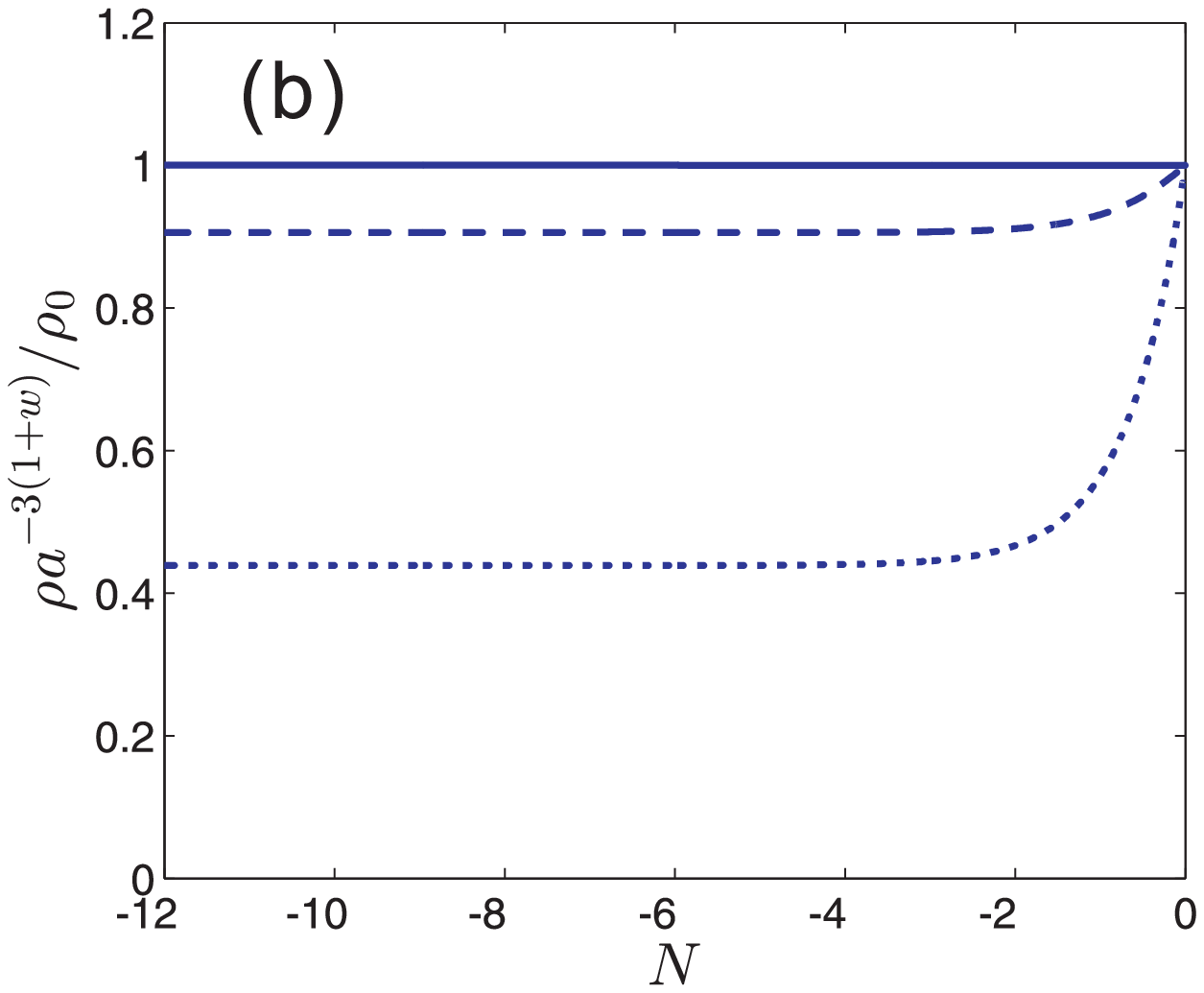}
\caption{Evolutions of $\rho_m a^3$ (blue lines) and $\rho_r a^4$ (green lines) with (a) $C_r = 1$ and (b) $C_r = 0$, where legend is the same as Fig.~\ref{fg:1}.}
\label{fg:5}
\end{figure}
Substituting Eq.~(\ref{eq:rhol}) into the conservation equation $\nabla^{\mu} (T^M_{\mu \nu}+T^\Lambda_{\mu \nu}) = 0$, we have
\begin{eqnarray}
\label{eq:contl}
\dot{\rho}_{\Lambda} + 3 H (1+w_{\Lambda}) \rho_{\Lambda} = \dot{\rho}_{\Lambda} \propto \dot{H} \neq 0 \,,
\end{eqnarray}
resulting in  that dark energy unavoidably couples to matter and radiation, given by
\begin{eqnarray}
\label{eq:contm}
&& \dot{\rho}_m + 3 H \rho_m = Q_m \,, \\
\label{eq:contr}
&& \dot{\rho}_r + 4 H \rho_r = Q_r \,,
\end{eqnarray}
where $Q_{m,r}$ are the decay rate from $\Lambda(t)$ to matter and radiation,  taken to be
\begin{eqnarray}
\label{eq:Qmr}
Q_i=\frac{\dot{\rho}_{\Lambda}C_i(\rho_i+P_i)}{\sum_{j=m,r} C_j(\rho_j+P_j)} \,,
\end{eqnarray}
respectively.
Note that the analytical solution of Eq.~(\ref{eq:contl}) has been obtained with $\lambda_0=0$ and $w_M=constant$ in Refs.~\cite{Borges:2005qs, Borges:2007bh}.
However, if $\lambda_0 \neq 0$ and $\rho_M = \rho_m + \rho_r$, composited of multi-fluid with EoS $w_r \neq w_m$, the analytical solution no longer exists.

In Fig.~\ref{fg:1}, we show the cosmological evolutions of $\rho_m$ (blue line), $\rho_r$ (green line) and $\rho_{\Lambda}$ (red line), normalized by the characteristic energy density $\rho_{ch}$, as functions of the e-folding $N \equiv \ln a$ with $\rho_m a^{3}/\rho_{ch} = 1$ and $\rho_r a^{4}/\rho_{ch} = 3 \times 10^{-4}$ at $N = -12$, where $(C_r,C_m)$ are (a) $(1,1)$, (b) $(1,0)$ and (c) $(0,1)$ with $(\lambda_0,\lambda_1)=(1,0)$, $(0.9,0.1)$ and $(0,1)$, corresponding to the solid, dashed and dotted lines, respectively.
In Fig.~\ref{fg:1}c, we observe that if dark energy fully decays to radiation, $\rho_r$ can be the same order of $\rho_m$ at $\lambda_1 \gtrsim 0.1$, which violates the current observations.
On the contrary, this problem never occurs if $\Lambda(t)$ only couples to matter as shown in Fig.~\ref{fg:1}b.
This behavior allows us to fix $C_m=1$ and keep $C_r$ to be a free parameter in the later study.
In Fig.~\ref{fg:5}, we present $a^3 \rho_m$ (blue line) and $a^4 \rho_r$ (green line) as functions of $N$ with $(\lambda_0,\lambda_1)=(1,0)$, $(0.9,0.1)$ and $(0,1)$, where the boundary conditions are the same as those in Fig.~\ref{fg:1}.
From the plots in Figs.~\ref{fg:1} and \ref{fg:5}, one can see that the cosmological constant decouples to matter and radiation (corresponding to $\rho_i a^{-3(1+w_i)}  = constant$) when $(\lambda_0, \lambda_1) \rightarrow (1,0)$.
On the other hand, a large number of matter and radiation are created by the decay of dark energy at the late-time of the universe with $(\lambda_0, \lambda_1) \rightarrow (0,1)$.
When dark energy dominates the universe, $\rho_\Lambda$ decays slowly and our universe turns into the accelerating expansion phase.
Clearly, this scenario is suitable to describe the dark energy problem no matter what values for $\lambda_0$ and $\lambda_1$ are used.
To examine this model in detail, we need to study the linear perturbation theory of gravity and investigate the effects at the subhorizon scale.

\section{Cosmological perturbation in running cosmological constant model}
\label{sec:perturbation}

We focus on the perturbation theory with the synchronous gauge  and explore the power spectra of matter and the CMB temperature
in the running cosmological constant model.
Under the FLRW background, the metric perturbation is given by~\cite{Ma:1995ey}
\begin{eqnarray}
\label{eq:pert_metric}
ds^2 = a^2(\tau) \left[ -d\tau^2 + (\delta_{ij} + h_{ij}) dx^i dx^j \right] \,,
\end{eqnarray}
where
\begin{eqnarray}
\label{eq:pert_h}
h_{ij} = \int d^3 k e^{i \vec{k} \cdot \vec{x}} \left[ \hat{k}_i \hat{k}_j h(\vec{k},\tau) + 6 \left( \hat{k}_i \hat{k}_j -\frac{1}{3} \delta_{ij} \right) \eta(\vec{k},\tau) \right] \,,
\end{eqnarray}
$i,j=1,2,3$, $h$ and $\eta$ are two scalar perturbations in the synchronous gauge, and $\hat{k} = \vec{k}/ k $ is the k-space unit vector. 
The matter (baryon, cold dark matter and massive neutrino) and radiation (photon and massless neutrino) density perturbations can be derived from the conservation equation $\nabla^{\mu} (T^M_{\mu \nu}+T^\Lambda_{\mu \nu}) = 0$ with $\delta
 T^0_0 = \delta \rho_M$, $\delta
 T^0_i = -T^i_0 = (\rho_M+P_M) v^i_M$ and $\delta
 T^i_j = \delta P_M \delta^i_j$, 
 given by~\cite{Arcuri:1993pb, Borges:2008ii},
\begin{eqnarray}
\label{eq:drho}
\dot{\delta}_M = - (1+w_M)\left( \theta_M + \frac{\dot{h}}{2} \right) - 3 H \left( \frac{\delta P_M}{\delta \rho_M} - w_M \right) \delta_M -\frac{Q_M}{\rho_M} \delta_M \,,
\end{eqnarray}
where $\delta_M \equiv \delta \rho_M / \rho_M$, $\theta_M = ik_i v^i_M$,
 and $Q_M$ is the decay rate in Eq.~(\ref{eq:Qmr}).

\begin{figure}
\centering
\includegraphics[width=0.49 \linewidth]{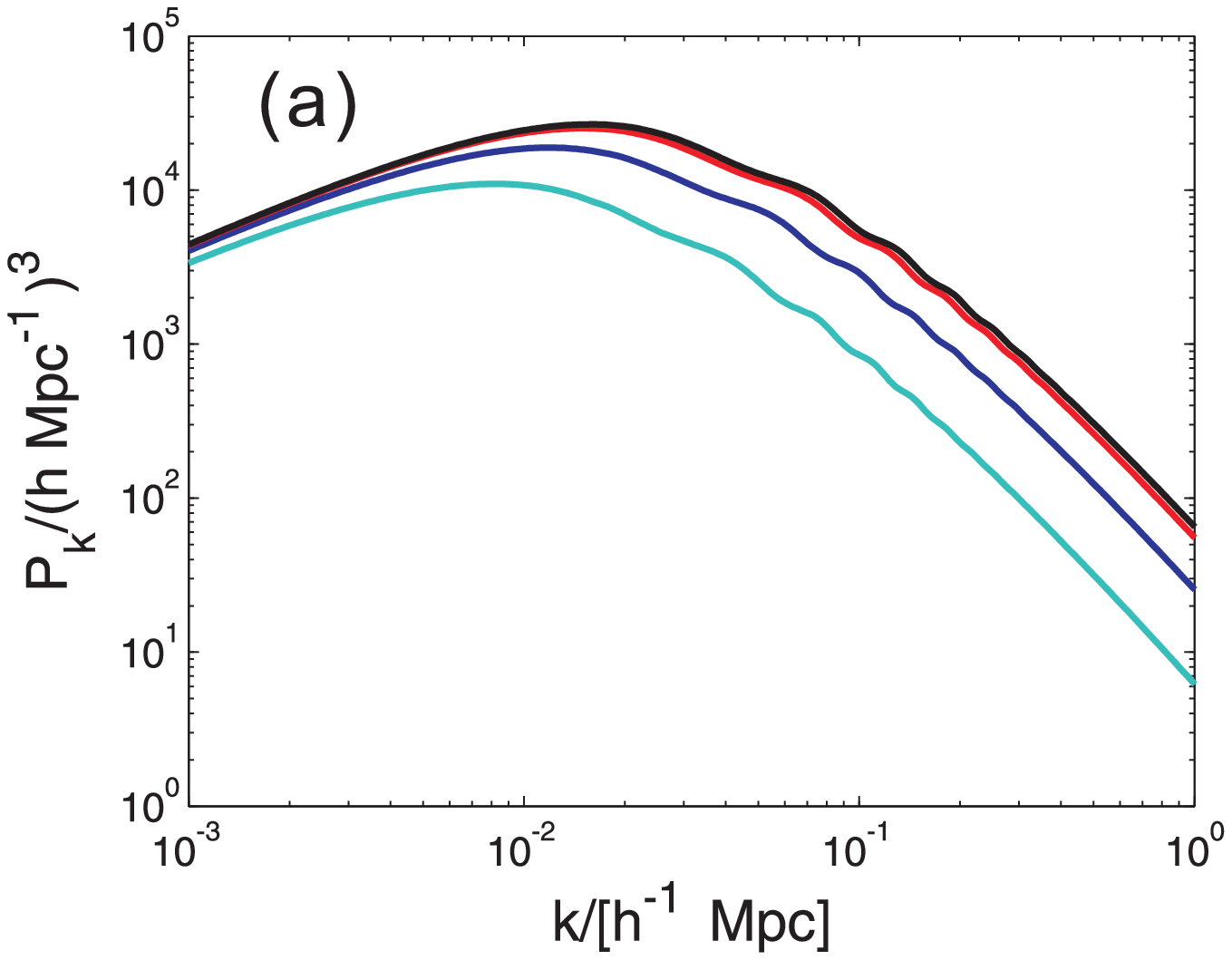}
\includegraphics[width=0.49 \linewidth]{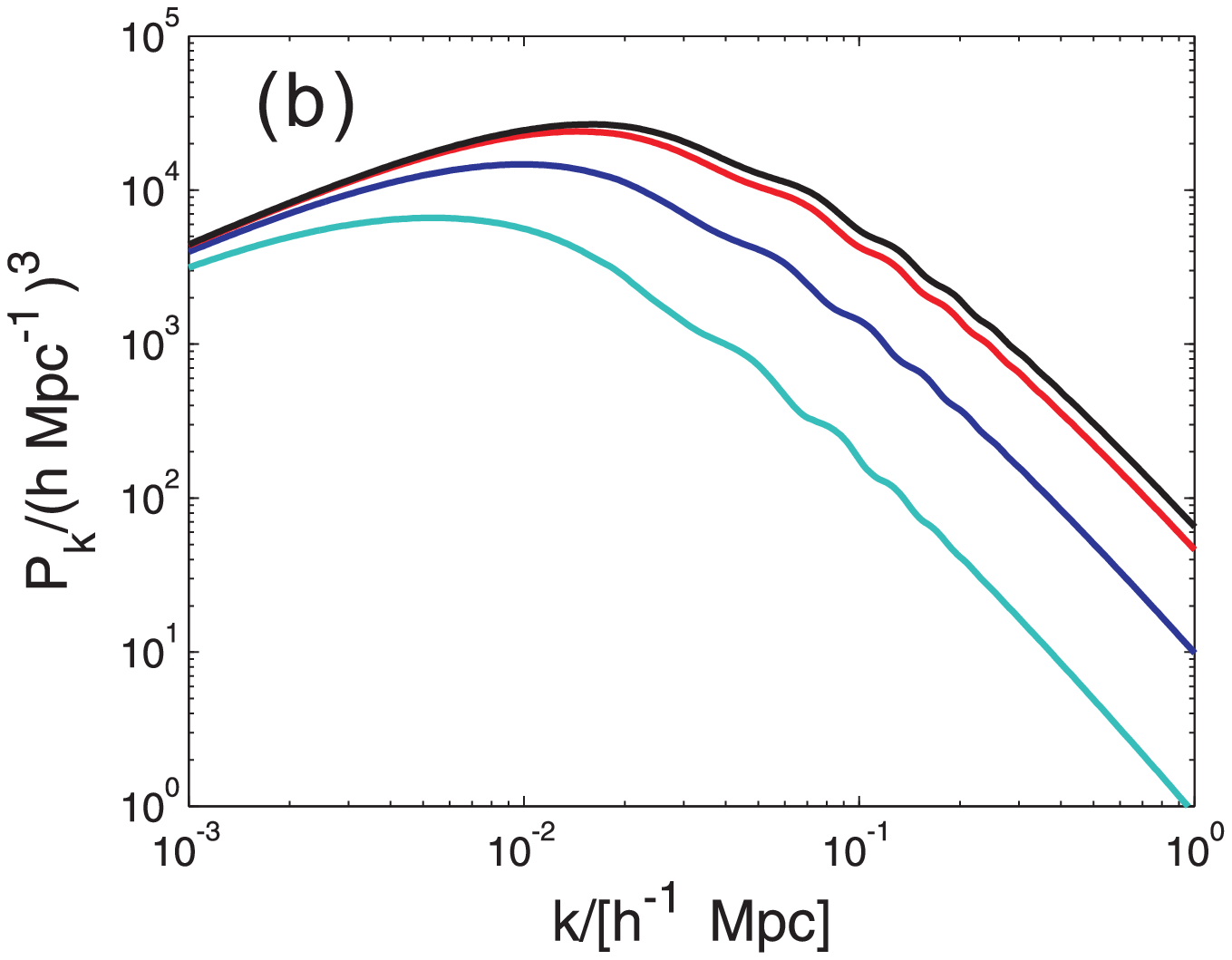}
\caption{
The matter power spectrum $P(k)$ as a function of the wavelength $k=2 \pi/\lambda$ with (a) $C_r = 1$ and (b) $C_r = 0$, where $(\lambda_0, \lambda_1) = (1, 0)$ (black line), $(0.9, 0.1)$ (red line), $(0.5, 0.5)$ (blue line), $(0, 1)$ (cyan line), and the boundary conditions are taken to be $\Omega_b h^2=2.23\times 10^{-2}$, $\Omega_c h^2=0.119$, $h=0.7$ and $ \Sigma m_{\nu}=0.06$~eV, respectively.
}
\label{fg:2}
\end{figure}
To show how the running cosmological constant scenario influences the matter density perturbation and CMB temperature fluctuation, we perform the open-source program {\bf CAMB}~\cite{Lewis:1999bs} with the model in Eq.~(\ref{eq:rhol}), and modify the evolution of the density perturbation in Eq.~(\ref{eq:drho}) as well as the background density evolutions $\rho_\Lambda(z)$, $\rho_m(z)$ and $\rho_r(z)$, solved from Eqs.~(\ref{eq:contl})-(\ref{eq:contr}), respectively.
 Since the dark energy density ratio, $\rho_{\Lambda}/\rho_c \propto H^{-1}$, is negligible in the early universe, 
 the running cosmological constant model shares the same initial condition as the $\Lambda$CDM model 
 with $\rho_M=\rho_M \lvert_{a \rightarrow 0}$.
Besides, the matter-radiation equality $z_{eq}$ also changes in this model, given by
\begin{eqnarray}
\frac{\rho_m(z=z_{eq})}{\rho_r(z=z_{eq})} = 1 \,.
\end{eqnarray}
In Fig.~\ref{fg:2}, we present the matter power spectrum $P(k)$ as a function of the wavenumber $k$ with (a) $C_r = 1$ and (b) $C_r = 0$, along with $\Omega_b h^2=2.23\times 10^{-2}$, $\Omega_c h^2=0.119$, $\Sigma  m_{\nu}=0.06$~eV and $(\lambda_0, \lambda_1) = (1, 0)$ (black line), $(0.9, 0.1)$ (red line), $(0.5, 0.5)$ (blue line) and $(0, 1)$ (cyan line), respectively.
Obviously, the matter power spectrum $P(k)$ is suppressed at a large $k$, and the deviation from $\Lambda$CDM (black line) is significant for $\lambda_1 \gg 0$.
As a result, the allowed parameter space for $\lambda_1$ can be estimated to be small.
The suppression of $P(k)$ can be realized from Eq.~(\ref{eq:drho}) with $Q_m \propto -\dot{H} \geq 0$.
As $\lambda_1$ increases, the running cosmological model deviates from the $\Lambda$CDM limit, while the matter creation gets enhanced due to the dark energy decay as seen in Eq.~(\ref{eq:contm}).
In addition, since dark energy is considered to be homogeneous and isotropic, $\delta \rho_\Lambda = 0$, so that the creation of matter smoothly distributes to our universe.
Consequently, the density perturbation is diluted and the matter power spectrum is suppressed by the decay of $\Lambda$.

\begin{figure}
\centering
\includegraphics[width=0.49 \linewidth]{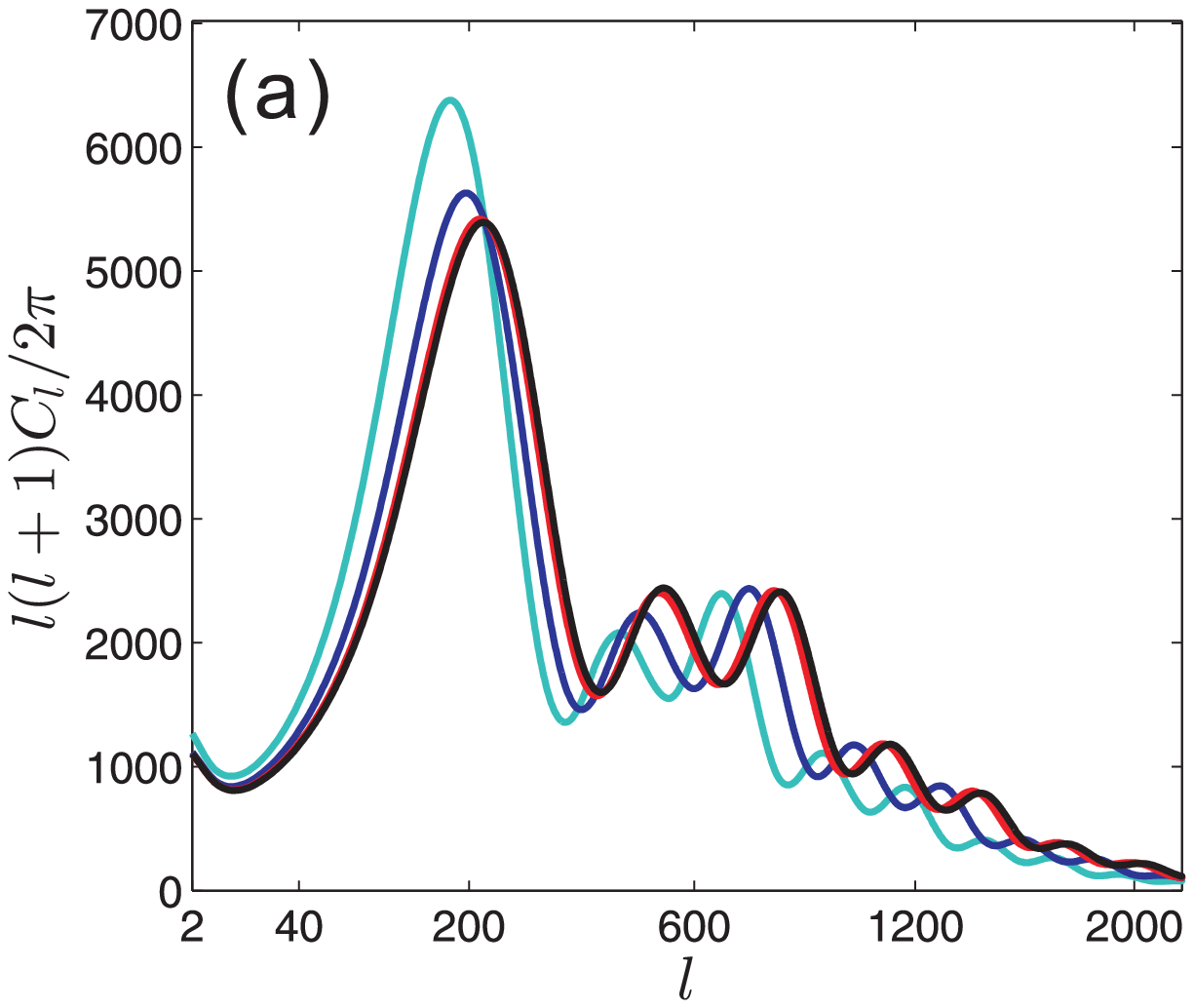}
\includegraphics[width=0.49 \linewidth]{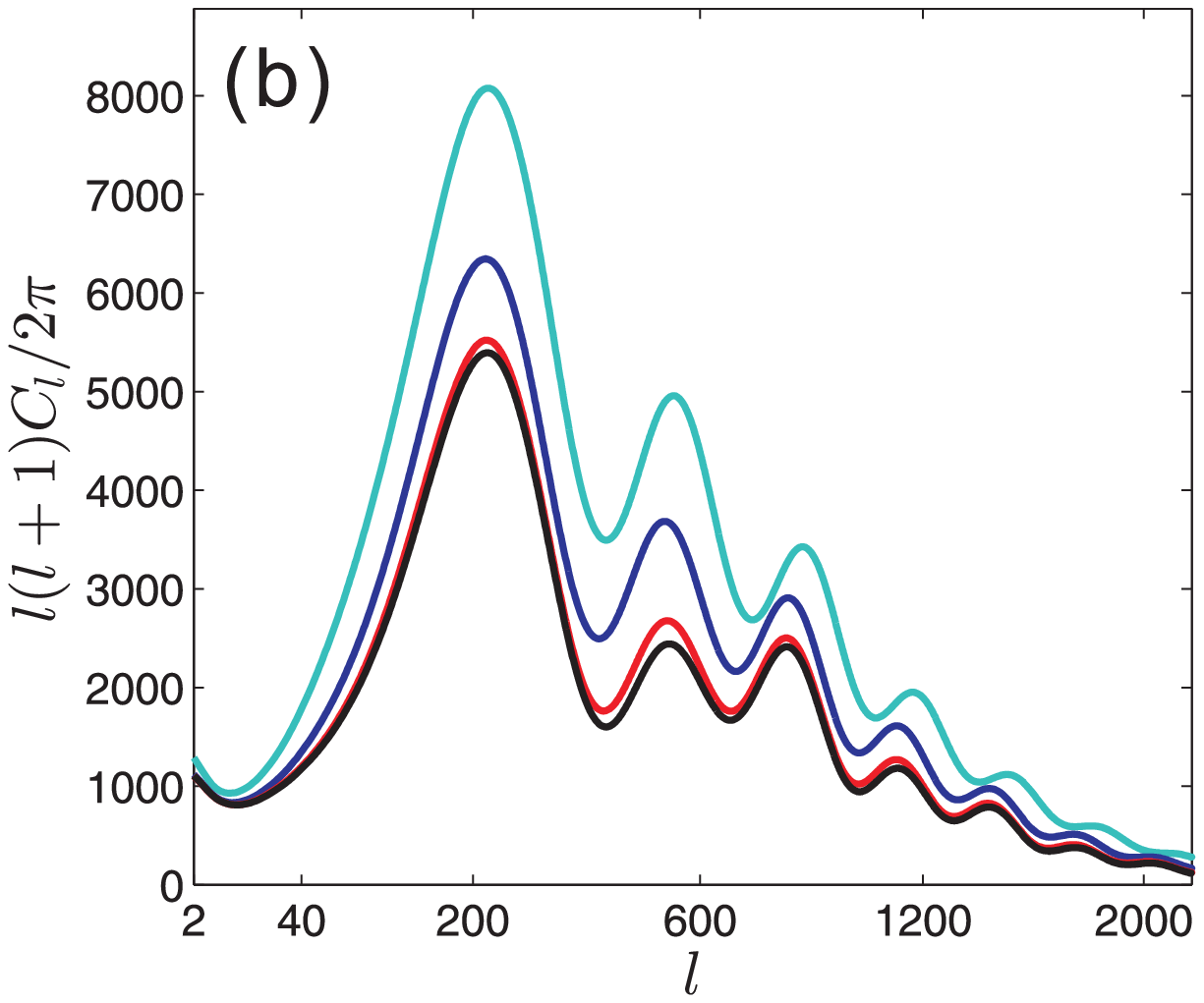}
\caption{
The CMB temperature power spectrum for (a) $C_r = 1$ and (b) $C_r = 0$ with $T=2.73~K$, where legend is the same as Fig.~\ref{fg:2}.
}
\label{fg:3}
\end{figure}
Fig.~\ref{fg:3} depicts  the CMB temperature power spectrum in the running cosmological constant model with (a) $C_r = 1$ and (b) $C_r = 0$, while the boundary conditions are the same as those in Fig.~\ref{fg:2}.
One finds that the CMB temperature spectrum in $\Lambda(t)$CDM significantly differs from that in  $\Lambda$CDM (black line)
when $\lambda_1 \gg 0$.
This gives us a hint that $\lambda_1$ would be relatively small in order to fit the spectrum.
Our results demonstrate that the allowed matter creation from the cosmological constant should be tiny.

\section{Observational constraints}
\label{sec:constraints}

\begin{table}[ht]
\begin{center}
\caption{Priors for cosmological parameters with $\rho_{\Lambda} = \rho_{\Lambda}^0 \left[ \lambda_0 + \lambda_1 H/H_0 \right]$ and $\lambda_0+\lambda_1=1$.  }
\begin{tabular}{|c||c|} \hline
Parameter & Prior
\\ \hline
Baryon density & $0.5 \leq 100\Omega_bh^2 \leq 10$
\\ \hline
CDM density & $10^{-3} \leq \Omega_ch^2 \leq 0.99$
\\ \hline
Optical depth & $0.01 \leq \tau \leq 0.8$
\\ \hline
Neutrino mass sum& $0 \leq \Sigma m_{\nu} \leq 2$~eV
\\ \hline
Model parameter $\lambda_1$& $0 \leq \lambda_1 \leq 0.3$
\\ \hline
\end{tabular}
%\vskip 0.2in
\label{tab:1}
\end{center}
\end{table}

\begin{figure}
\centering
\includegraphics[width=0.98 \linewidth]{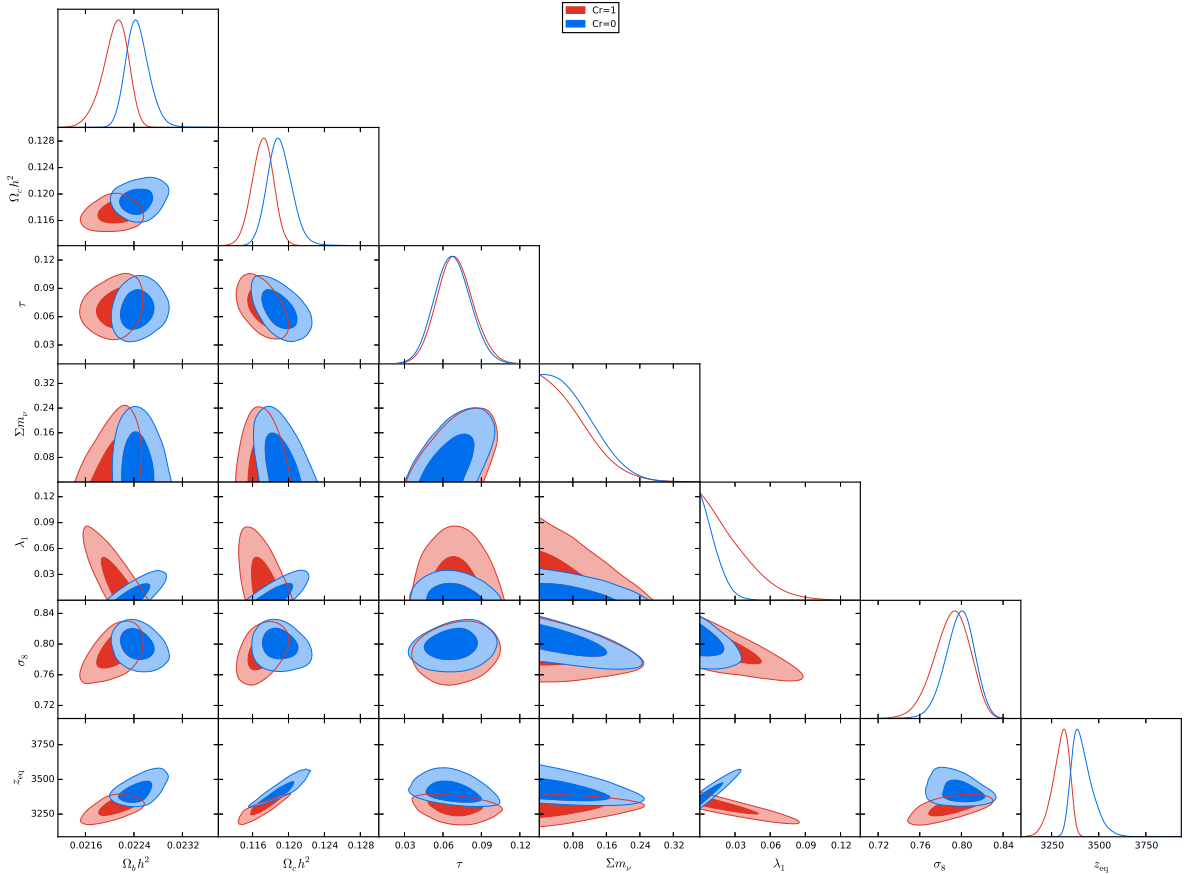}
\caption{
One and two-dimensional distributions of  $\Omega_bh^2$, $\Omega_ch^2$, $\tau$,  $\Sigma m_{\nu}$, $\lambda_1$, $\sigma_8$ and  $z_{eq}$, where the contour lines represent  68\% and 95\% confidence levels, respectively. }
\label{fg:4}
\end{figure}

We now examine the possible ranges for $\lambda_0$ and $\lambda_1$ by the cosmological observations.
We use the open-source {\bf CosmoMC} program~\cite{Lewis:2002ah} with the MCMC method to perform the global fitting for the $\Lambda(t)$CDM model.
The dataset includes (i) {\it Planck 2015} TT, TE, EE, low-$l$ polarization and lensing from SMICA~\cite{Adam:2015wua,Aghanim:2015xee,Ade:2015zua}; (ii) baryon acoustic oscillation (BAO) data from 6dF Galaxy Survey~\cite{Beutler:2011hx}, SDSS DR7~\cite{Ross:2014qpa} and BOSS~\cite{Anderson:2013zyy}; (iii) matter power spectrum data from SDSS DR4 and WiggleZ~\cite{Blake:2011en}, and (iv) weak lensing data from CFHTLenS~\cite{Heymans:2013fya}.
With this dataset, we
keep the model parameter $\lambda_1 \geq 0$ to avoid the negative dark energy density and explore the constraints on $\lambda_1$
and other cosmological parameters.
The priors of the various parameters are listed in Table.~\ref{tab:1}, and our results are shown in Fig.~\ref{fg:4}.

From Fig.~\ref{fg:4}, we find that 
 $\lambda_1$ is strongly constrained with $\lambda_1 = \sigma H_0/(\sigma H_0+\Lambda_0) \simeq \sigma H_0/\Lambda_0 \lesssim 6.68 \times 10^{-2}$ ($2.63 \times 10^{-2}$)
for $C_r = 1 (0)$ at $2\sigma$-confidence level, which is consistent with the prediction in Sec.~\ref{sec:perturbation}.
Additionally, the best-fit $\chi^2_{\Lambda(t)} = 13546.5$ ($13545.7$) for $C_r = 1 (0)$ in the $\Lambda(t)$CDM is slightly larger than
$\chi^2_{\Lambda} = 13545.3$  in the $\Lambda$CDM model.
Since the cosmological observations prefer the $\Lambda$CDM limit, the effect from the decaying dark energy model is strongly confined.
% in this model.
Finally, the allowed ranges for the various cosmological parameters with $2\sigma$-confidence level are summarized in Table.~\ref{tab:2}.

\begin{table}[ht]
\begin{center}
\caption{List of allowed regions as $95\%$ C.L. with $\rho_{\Lambda} = \rho_{\Lambda}^0 \left[ \lambda_0 + \lambda_1 H/H_0 \right]$ and $\lambda_0+\lambda_1=1$,
where $\chi^2_{\mathrm{Best-fit}}=\chi^2_{\mathrm{CMB}}+\chi^2_{\mathrm{BAO}}+\chi^2_{\mathrm{MPK}}+\chi^2_{\mathrm{lensing}}$.} 
\begin{tabular}{|c||c|c|c|} \hline
Parameter & $\Lambda(t)$CDM with $C_r=1$ & $\Lambda(t)$CDM with $C_r=0$ & $\Lambda$CDM
\\ \hline
Model parameter ($\lambda_1$) & $<6.68 \times 10^{-2}$ & $<2.63 \times 10^{-2}$ & $ 0$
\\ \hline
Baryon density ($ 100 \Omega_bh^2$) & $ 2.21 ^{+0.04}_{- 0.05}$  & $ 2.25 \pm 0.04$ & $ 2.23 \pm 0.04$
\\ \hline
CDM density ($ \Omega_ch^2 $) & $ 0.117 \pm 0.002$ & $0.119^{+0.003}_{-0.002}$ & $ 0.118 \pm 0.002$
\\ \hline
Optical depth ($ \tau$) & $  6.71^{+2.83}_{-2.62} \times 10^{-2} $ & $ 6.74^{+2.79}_{-2.62} \times 10^{-2} $ & $ 6.99^{+2.83}_{-2.77}\times 10^{-2} $
\\ \hline
Neutrino mass sum ($\Sigma m_{\nu} $) & $< 0.189$~eV & $ < 0.194$~eV & $< 0.221$~eV
\\ \hline
$\sigma_8$ & $ 0.791^{+0.034}_{-0.035}$  & $ 0.799^{+0.025}_{-0.029}$ & $  = 0.805^{+0.026}_{-0.028}$
\\ \hline
$z_{eq}$  & $3298^{+82}_{-94} $ & $ 3414^{+117}_{-98} $ & $3351 \pm 46 $
\\ \hline
 $\chi^2_{\mathrm{Best-fit}}$  & $13546.5$ & $ 13545.7$ & $ 13545.3$
\\ \hline
 $\chi^2_{\mathrm{CMB}}$  & $13032.4$ & $ 13032.2$ & $ 13031.1$
\\ \hline
$\chi^2_{\mathrm{BAO}}$  & $4.3$ & $ 4.3$ & $ 4.8$
\\ \hline
$\chi^2_{\mathrm{MPK}}$  & $479.9$ & $ 479.6$ & $ 480.0$
\\ \hline
$\chi^2_{\mathrm{lensing}}$  & $29.9$ & $ 29.5$ & $ 29.4$
\\ \hline
\end{tabular}
%\vskip 0.2in
\label{tab:2}
\end{center}
\end{table}

\section{Conclusions}
\label{sec:conclusion}

We have investigated the $\Lambda(t)$CDM model with the dark energy decaying to both matter and radiation, in which $\Lambda(t)= \sigma H + \Lambda_0$.
Although this scenario is suitable to describe the late-time accelerating universe at the background level, the linear perturbation analyses of the matter power and CMB temperature spectra have set a strong constraint on the model parameter $\lambda_1$ in Eq.~(\ref{eq:rhol}).
Explicitly, by performing the global fit from the observational data, we have obtained that 
$\lambda_1 \simeq \sigma H_0/\Lambda_0 \lesssim 6.68 \times 10^{-2}$ ($2.63 \times 10^{-2}$) and 
$\chi^2_{\Lambda(t)\mathrm{CDM}}=13546.5 (13545.7) 
 \gtrsim \chi^2_{\Lambda\mathrm{CDM}}=13545.3$ for $C_r = 1 (0)$, implying that the current data prefers the $\Lambda$CDM limit.
Constraints on other cosmological parameters in both $\Lambda(t)$CDM and $\Lambda$CDM models have been also given in 
Table~\ref{tab:2}.

\section*{Acknowledgments}
This work was partially supported by National Center for Theoretical Sciences, National Science Council (NSC-101-2112-M-007-006-MY3), MoST (MoST-104-2112-M-007-003-MY3), and
 National Natural Science Foundation of China  (11447104 and  11505004).
%  and National Tsing Hua University~(104N2724E1).

\end{document}